\documentclass[final,1p, times,twocolumn]{elsarticle}
\usepackage{amssymb}
\usepackage{lineno}
\usepackage{xcolor}
\journal{Nuclear Instruments and Methods A}
\begin{document}
\begin{frontmatter}
\title{A Tetramethylsilane TPC with Cherenkov light readout and 3D reconstruction}
\author[stanford]{S.X.~Wu}
\ead{sxwu@stanford.edu}
\author[stanford]{B.G.~Lenardo}
\ead{blenardo@stanford.edu}
\author[stanford]{M.~Weber\footnote{Now at Descartes Labs}}
\author[stanford]{G.~Gratta} 
\address[stanford]{Physics Department, Stanford University, 382 Via Pueblo Mall, Stanford, CA 94305, USA}

\begin{abstract}
We describe the construction and calibration of a multi-channel liquid time projection chamber filled with Tetramethylsilane. Its charge readout system consists of 8 wires each in the $X$ and $Y$ directions. The chamber is also equipped with a light readout system consisting of a 5'' photomultiplier tube coupled to the liquid volume through a viewport. The energy scale of the detector is calibrated using positron-electron pairs produced by 4.4 MeV gamma rays emitted from an AmBe source, using an external trigger on the positron annihilation gammas. The external trigger is then reconfigured to tag cosmic ray muons passing through the active Tetramethylsilane volume, which are used to measure the stopping power in Tetramethylsilane and the electron lifetime in the detector. We find a most-probable energy loss from minimum ionising particles of $\Delta_p/ds$ = (0.60$\pm$0.01) MeV/cm. We also measure an electron lifetime of 43$^{+680}_{-21}~\mu$s by measuring the most-probable energy loss as a function of drift time. For both measurements, the errors are statistical only. For both fast electron and muon signals, the photomultiplier tube detects prompt Cherenkov light, demonstrating the possibility of self-triggering of the detector. The room-temperature organic target medium, together with the self-triggering capabilities and long electron lifetimes reported in this work, make this an attractive technology to explore for rare event detectors or other applications in the area of radiation measurements. 
\end{abstract}
\begin{keyword}
Ionisation detector \sep Organic liquid \sep Tracking and calorimeter
\end{keyword}
\end{frontmatter}

\section{Introduction}
\label{S:1}
Ionization detectors utilizing room temperature organic liquids (LOr) have been explored in the 1980's and 1990's for calorimetry in high energy physics~\cite{UA1_TMP_Calo, engler_1999}, but little recent work exists in the literature. Meanwhile, tremendous progress has been achieved in the intervening decades using cryogenic liquefied noble elements (mainly liquid xenon (LXe) and liquid argon (LAr)) in radiation detectors. Perhaps most important is relatively recent work demonstrating the ability to drift electrons over large distances, along with various schemes to combine the signal from the prompt scintillation light with that from the ionization directly collected in the detector.  These innovations enabled the construction of large-target-mass Time Projection Chambers (TPCs), as opposed to simpler ionization detectors, and opened the way to their use in situations where the event time is not externally provided (e.g. by an accelerator). Noble liquid TPCs have found many applications in fundamental physics (see, e.g. Ref.~\cite{EXO2002019_CompleteAnalysis,LUX2017_CompleteExposure,XENON1T2018_1TonYearResults}), and are also being explored for applications in radiation imaging~\cite{Grignon2007_LXePET,Aprile2008_LXeComptonImaging,Nishikido2004_LXePET,Wahl2012_PiXeYComptonImaging,GallegoManzano2015_LXePET}.

Organic liquids offer a variety of trade-offs with respect to LXe and LAr.  They are generally in liquid phase at or near room temperature, resulting in very substantial simplification of the apparatus. This advantage may be particularly important for industrial or medical applications where simplicity of operation and ruggedness are paramount, or in applications where the radiation to be detected is produced outside of the TPC, as the inactive material can be minimized. Furthermore, molecular liquids offer the possibility, at least in principle, of choosing the elements that they contain, broadening the array of nuclear properties that can be exploited.  As an example, most organic materials contain substantial amounts of hydrogen, which could enable the detection of antineutrinos via the inverse beta decay reaction ($\bar{\nu_e} + p \rightarrow e^+ + n$)~\cite{Dawson_2014}. In this work we use tetramethylsilane (Si[CH$_3$]$_4$), but other materials obtained by the substitution of a Si atom with Ge, Sn or Pb are known to have similar properties from the electron drift point of view, providing access to heavy elements~\cite{Geer1991}. Finally, it is plausible that even a single material, such as Tetramethylsilane (TMS), could be doped in a way similar to liquid scintillators~\cite{Dayabay_LSC}, to enable new capabilities for specific applications. Due to these advantages, there are ongoing efforts to develop trimethylbismuth (TMBi) detectors for next-generation positron-emission tomography~\cite{Farradeche_2018}, as well as efforts to drift charge through LOr scintillator media for use in neutrino experiments~\cite{MCCONKEY2013459}.

Among the drawbacks, LOr media are often flammable or hazardous and they are more difficult to obtain in a very pure state. Rare event searches may suffer from $^{14}$C backgrounds at low energies, though this may be suppressed by using underground sources of carbon. In many cases, LOr media used in ionization chambers are also unlikely to emit scintillation light in a frequency band where common photodetectors are sensitive. Finally, at least for the case of TMS and tetramethylpentane (TMP), commonly available electric fields are far from achieving full saturation of the electron collection~\cite{Engler_1996} so that the charge yield is modest with respect to LXe and LAr.  While this certainly limits the energy resolution achievable, its origin, related to differences in the Onsager radius and the physics of recombination~\cite{Onsager_1938} may offer some complementarity to LXe or LAr in terms of details of the interactions of different types of particles. 

The relative ease with which modern LXe and LAr detectors obtain long electron drift lengths is generally due to advances in vacuum technology (dry pumps) and purification systems, and great care in selecting the materials in contact with the liquids. There is the hope that some of those techniques and the knowledge gathered from the work on LXe and LAr could be transferred to LOr detectors, making sizeable TPCs possible. In addition, the lack of scintillation light may be, at least in part, remedied by the detection of Cherenkov radiation.

In this paper we report on a new exploration of LOr detector media, with steps towards constructing proper TPCs. As a first step we have built a detector that is filled with TMS and operates at room temperature. TMS was chosen because of its non-polar structure and relatively high electron mobility and charge yield compared with other previously-explored LOr media~\cite{Engler_1996}.
The ionization readout is segmented by two perpendicular arrays of wires, in one case detecting charge by induction and in the other directly collecting it. The event time can be provided, to our knowledge for the first time, by the detection of Cherenkov light, enabling self-triggering of the detector.
This technique, common in LAr and LXe detectors, is used for a LOr ionization detector here for the first time. 
We perform calibration measurements using both fast electrons and cosmic ray muons, enabling a direct measurement of the electron lifetime in the liquid TMS. While a relatively short drift length of 15.5~mm is used in this first detector, 
our results suggest that significantly longer drift distances are easily achievable.

\section{TMS detector and support system}
\label{S:2}

Since liquid purity is a concern, great care is applied in avoiding, for all wetted components, materials that may leach or outgas in the TMS. Only metals (304 stainless steel (SS) and copper), ceramics and glass are used in the internal structure. Furthermore, employing well-established ultra-high vacuum (UHV) techniques, the detector is designed in such a way so that trapped gas volumes are eliminated.  

\subsection{The time projection chamber}
 
\begin{figure}[t!]
\centering
\includegraphics[width=0.9\linewidth]{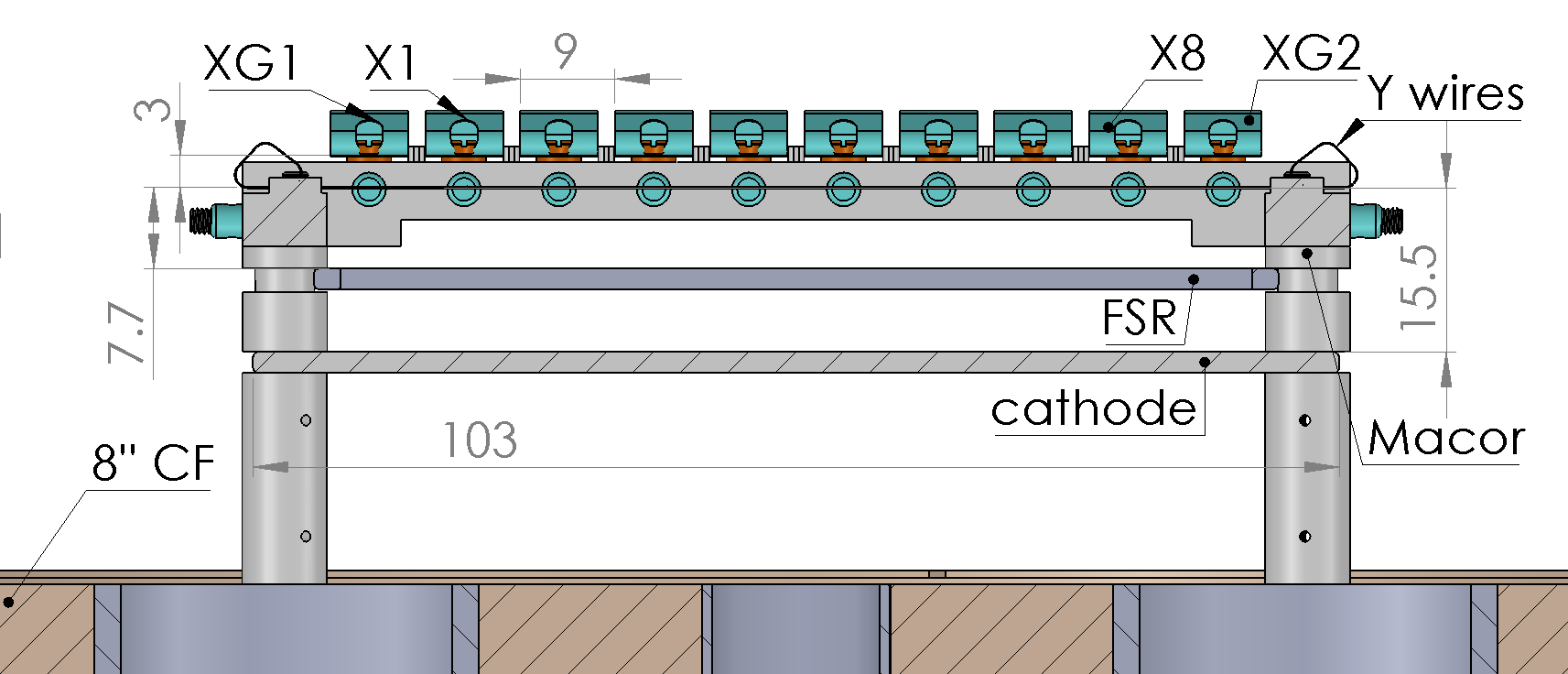}
\includegraphics[width=0.9\linewidth]{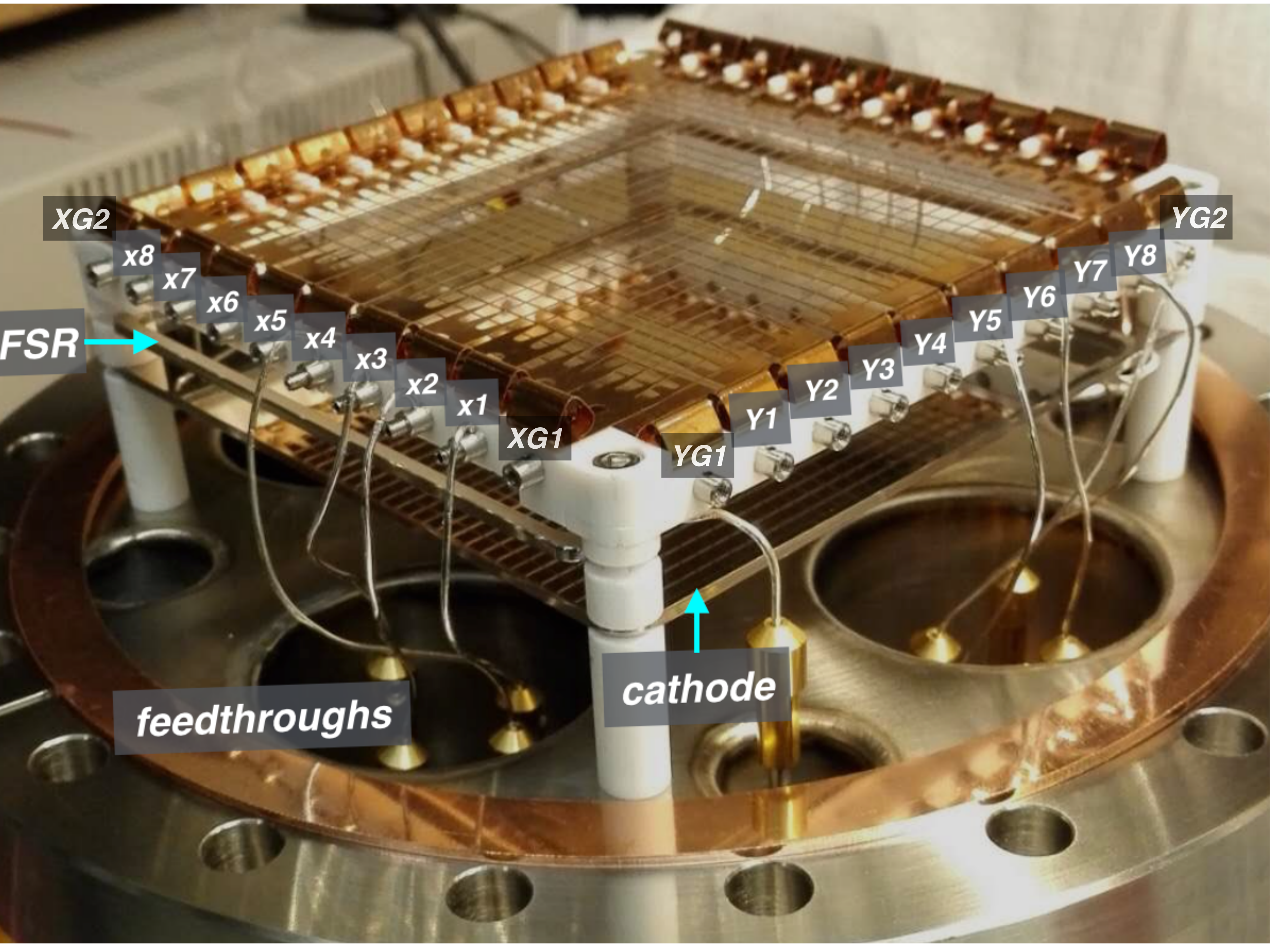}
\vspace{0.2cm} 
\caption{Top: schematic drawing of the TPC (dimensions in mm); bottom: photograph of the TPC structure.}
\label{fig:wire_chamber}
\end{figure}
 
 \begin{figure}[t!]
\centering
\includegraphics[width=0.37\linewidth]{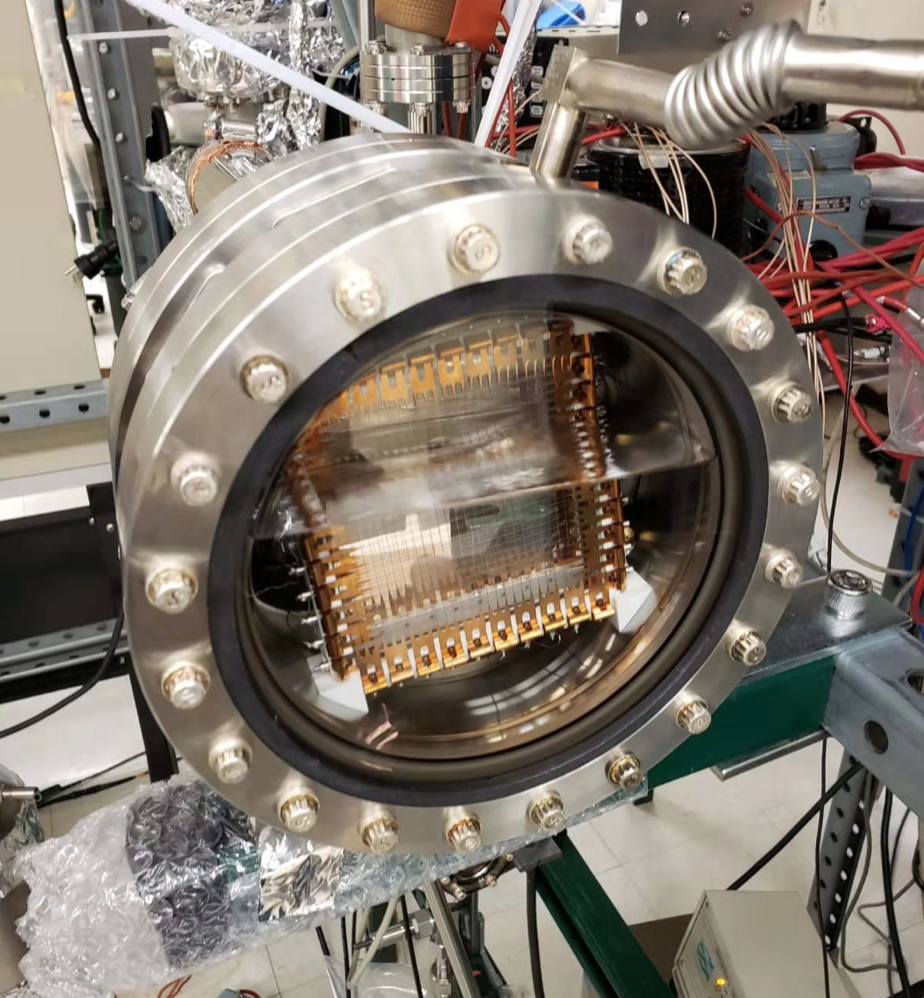} \hspace{0.05cm}
\includegraphics[width=0.52\linewidth]{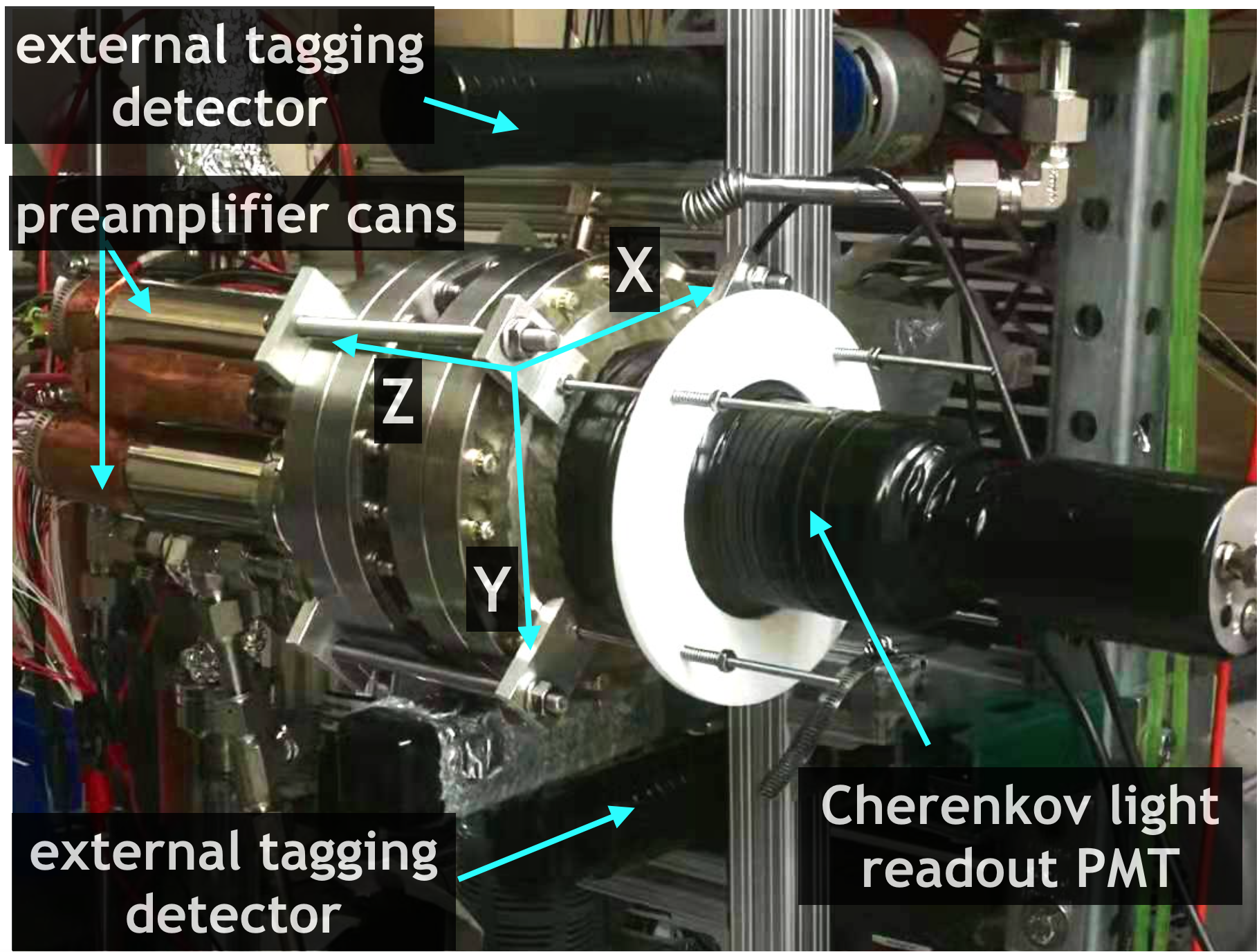}
\caption{Left: wire chamber seen through the viewport during TMS filling; right: overall experimental setup with Cherenkov photomultiplier tube and external tagging detectors.}
\label{fig:wire_chamber2}
\end{figure}
 
As shown in Fig.~\ref{fig:wire_chamber}, the TPC is constructed inside of a cylindrical vessel and mounted on an 8'' ConFlat\textsuperscript{\textregistered} (CF) flange. The readout consists of two planes of orthogonal wires -- eight in the $X$ direction and eight in the $Y$ direction -- supported by a frame structure machined from Macor\textsuperscript{\textregistered} glass ceramic. The cathode, a field shaping ring (FSR), and the wire planes are assembled with Macor spacers. The Macor frame also provides the connection of the wires to the feedthrough flanges where the readout electronics boards are located outside of the chamber. The cathode is a 2~mm thick, 103$\times$103~mm$^2$ SS plate. It is electro-polished and filleted at all four corners to minimise the risk of discharges to the grounded walls. The FSR is a 2~mm thick SS frame with a 105$\times$91~mm$^2$ outer and a 103$\times$89~mm$^2$ inner dimension and is used to improve the uniformity of the electric field over the fiducial volume. 

The readout wires are manufactured from phosphor bronze and photo-etched to a triplet pattern. The readout pitch is set to 9 mm, leading to triplets of wires with a 3 mm separation. This wire design follows the charge readout system of the EXO-200 experiment~\cite{Auger_2012}. The $X$ wires lay 3~mm above the $Y$ wires. The distance between the $Y$ wires and the cathode, which defines the electrons' maximal drift length, is 15.5~mm. Besides the eight $X$ and $Y$ wires reading out the charge signals, there is one additional wire placed at each end of the $X$ and $Y$ grids as shown in Fig.~\ref{fig:wire_chamber} ($XG1$,~$XG2$,~$YG1$ and $YG2$). The additional guarding wires are not instrumented, but are biased at the same potential as the $X$ and $Y$ wires to keep the electric field uniform in the sensitive volume of the detector. During operation, the electric field between the $X$ and $Y$ wires is kept at twice the field between the $Y$ wires and the cathode to guarantee a full electric transparency of the $Y$ wires~\cite{Aprile_1985}.
    
The 8-inch end flange is equipped with a 13.5~cm alkali borosilicate glass viewport. A 5-inch diameter hemispherical photomultiplier tube\footnote{ET Enterprises, model 9372B} is coupled to the flat window of the viewport via a plano-concave acrylic adapter using silicone optical grease\footnote{Bicron BC-630}, as seen in Fig.~\ref{fig:wire_chamber2}. The photomultiplier tube (PMT) is held at an operating bias of 1300V. At this voltage, single photoelectrons are well-separated from the noise. 
 
\subsection{Purification system}


\begin{figure}[t!]
\includegraphics[width=1\linewidth]{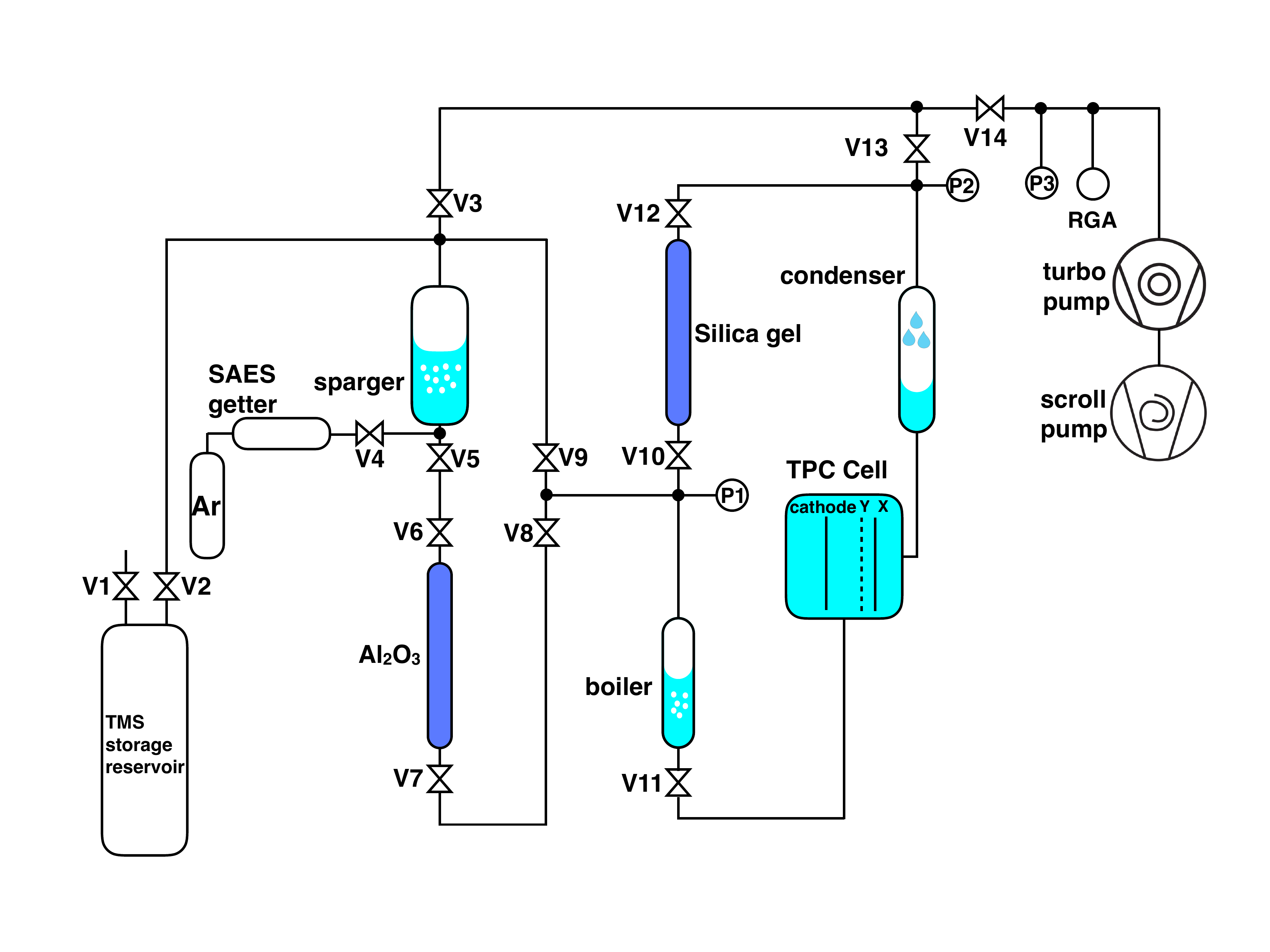}
\caption{Schematic of the TMS purification system. All metal valves are indicated with ``V", pressure gauges with ``P".}
\label{fig:pid}
\end{figure}

The TMS recirculation and purification system is schematically shown in Fig.~\ref{fig:pid}. It is designed to remove electronegative impurities, which can capture electrons as they drift across the detector volume and attenuate the measured ionization signal.\footnote{In liquid argon detectors, for example, the electron lifetime at an applied drift field of 1~kV/cm is related to the impurity level by the relationship $\tau_e \approx 500 \mu s/ \rho$, where $\rho$ is the impurity concentration in ppb oxygen-equivalent (meaning the impact of each species is weighted by its electronegativity relative to oxygen) \cite{Marchionni_2013_LArReview}. We expect a similar relationship to hold in LOr media.} Oxygen (O$_2$) and water (H$_2$O) are common electronegative impurities observed in noble liquid detectors, and we specifically target these two species.
All parts are UHV compatible, using all-metal seals in either CF or VCR  standards. Before filling the TMS, the system is pumped to a vacuum level better than $10^{-6}$~mbar. The TMS liquid\footnote{Sigma-Aldrich, $\geq$99.99\% purity, electronic grade} is initially stored inside a separate SS reservoir. 

During the filling, V2 is opened and the TMS is evaporated from the reservoir and re-condensed inside the SS sparging vessel by cooling the outside of the vessel with dry ice (V3, V5, and V9 remain closed). The liquid is then sparged using Ar gas\footnote{$\geq$99.9999\% purity, Praxair} that is pre-purified using a getter\footnote{SAES MicroTorr MC1-203F}. During sparging, the Ar is continuously pumped out through V3 and V14 with a scroll pump, efficiently removing the bulk of the oxygen and water vapor dissolved in the TMS. After sparging, V3 and V14 are closed and V5, V6, V7, and V8 are opened, allowing gravity to pull the TMS liquid through a column filled with $\sim$350 g aluminum oxide\footnote{Al$_{2}$O$_{3}$, Sigma Aldrich 199974} to further remove moisture. Finally, V5 and V6 are closed and V10 and V12 are opened, connecting the TMS liquid to the TPC recirculation system. To fill the TPC, the condenser is cooled to a temperature of 15$^o$\,C , reducing the pressure of the TMS vapor in the system. This allows the TMS to evaporate from the outlet of the aluminum oxide column, pass through the silica gel column (described below), and condense into the TPC. 

Once the TMS has been filled into the TPC, all valves except V10, V11, and V12 are closed to allow recirculation of the TMS in a closed loop during normal operation. A boiler, located downstream of the detector cell, consists of a recessed copper block and is maintained at a fixed temperature of 30$^{\circ}$C by means of an external cartridge heater driven by a proportional-integral-derivative (PID) controller. The TMS evaporated by the boiler passes through the column filled with $\sim$200 g silica gel\footnote{SiO$_{2}$, Sigma Aldrich 236802} which continuously removes residual moisture. Both the aluminum oxide and silica gel columns are activated by baking them under vacuum at 200$^{\circ}$C for four days before operation. The purified TMS vapor is condensed by a copper coil recessed in a ConFlat spool piece. Cooling water at a fixed temperature of 20$^{\circ}$C runs continuously inside the copper coil to provide cooling power. The condensed TMS liquid drips back into the detector cell. During data taking, the TMS liquid is continuously purified by this loop with a flow rate of $\sim$2~g/min, resulting in the total mass of 2.4~kg to be recirculated approximately every 20 hours. This recirculation is driven by the temperature difference between the boiler and condenser, with the temperature of the detector cell kept at 22$\pm0.5^{\circ}$C by the air-conditioning system in the laboratory. The first two filling stages are sufficient to remove most of the impurities, and charge signals are observed before the start of the continuous recirculation. To ensure the best performance, however, all measurements reported in this work were made after at least one week of continuous recirculation.

Extensive work has been done elsewhere on the extraction of heavy metals from organic liquid scintillators to reduce backgrounds in large-scale neutrino experiments~\cite{ford_snoplus,Borexino_purification}. While not employed in the present work, these techniques could be readily adapted in future applications of organic liquid TPCs.

\subsection{Readout electronics and data acquisition}

Wire channels are instrumented with charge-sensitive amplifiers made from discrete components, as described in Ref.~\cite{Jewell_2018}. The ionization arriving on the readout wires is fed to an RC integrator with a 500~$\mu$s time constant.
This is far larger than the maximum drift time of several $\mu$s, resulting in the full integration of the charge. An additional gain stage of two is placed after the charge integrator to better match the dynamic range of the digitizer. 
This architecture provides low noise characteristics at room temperature (500 $e^{-}$ RMS at a peaking time greater than 500~ns which is the case for the digital filters to be described in the following section for a several pF input capacitance of the readout wire), compact footprint, and the ability to drive cables over a long distance. The full functionality of the readout preamplifier is described in Ref.~\cite{Fabris_1999}.
 
The readout system is designed to allow both the $X$ and $Y$ wire planes to be independently biased at high voltages to enable the efficient drift and collection of ionization electrons. It relies on a signal decoupling stage implemented for each wire channel, consisting of a 220~pF, 5~kV capacitor and a 100~M$\Omega$ series resistor, which transmits the fast induction or collection signals to the preamplifiers. In this work, we ground the $Y$ wires and bias the $X$ wires at +3.2~kV. The FSR and cathode are biased at -4.4~kV and -8~kV, respectively.
 
The preamplifiers are installed in four SS cans visible in Fig.~\ref{fig:wire_chamber2} (right), mounted just outside the TPC feedthroughs to minimize the distance from the readout wires to the front-end electronics ($\sim$10~cm). For each can, an SHV connector is used to supply the bias to the $X$ channels, while signals are extracted via four LEMO connectors. Low voltage power to the preamps is also provided.
The SS hardware containing the TPC is grounded at the same potential as the HV ground and is insulated from the recirculation system by the boiler and condensor, which each includes a glass break in the plumbing.

The output signals from the 8+8 preamplifiers, along with signals from the Cherenkov light readout PMT and the external tagging detectors, are digitized at 62.5~MS/s with two Struck 3316 16-bit digitizers. 
A data acquisition window is set to 128~$\mu$s (8000 samples) with a 32~$\mu$s (2000 samples) window before the PMT trigger used for the baseline calculation. A drift field of 5.2~kV/cm is provided by the -8~kV cathode biasing.
At this field, the electron drift velocity is around 5.5~mm/$\mu$s taking into account the value of the mobility at 105 cm$^2$V$^{-1}$s$^{-1}$~\cite{Engler_1996}. The maximum drift distance of 15.5~mm corresponds to drift times $\leq$2.8~$\mu$s, well within the DAQ sampling window. For data taking with the AmBe source, the sampling rate is set at 125~MS/s for better timing resolution. In this case, the data-taking window contains 4000 samples (32~$\mu$s) with a 1000-sample pretrigger window (8~$\mu$s).

\subsection{Waveform analysis}

\begin{figure}[ht!]
\includegraphics[width=0.95\linewidth]{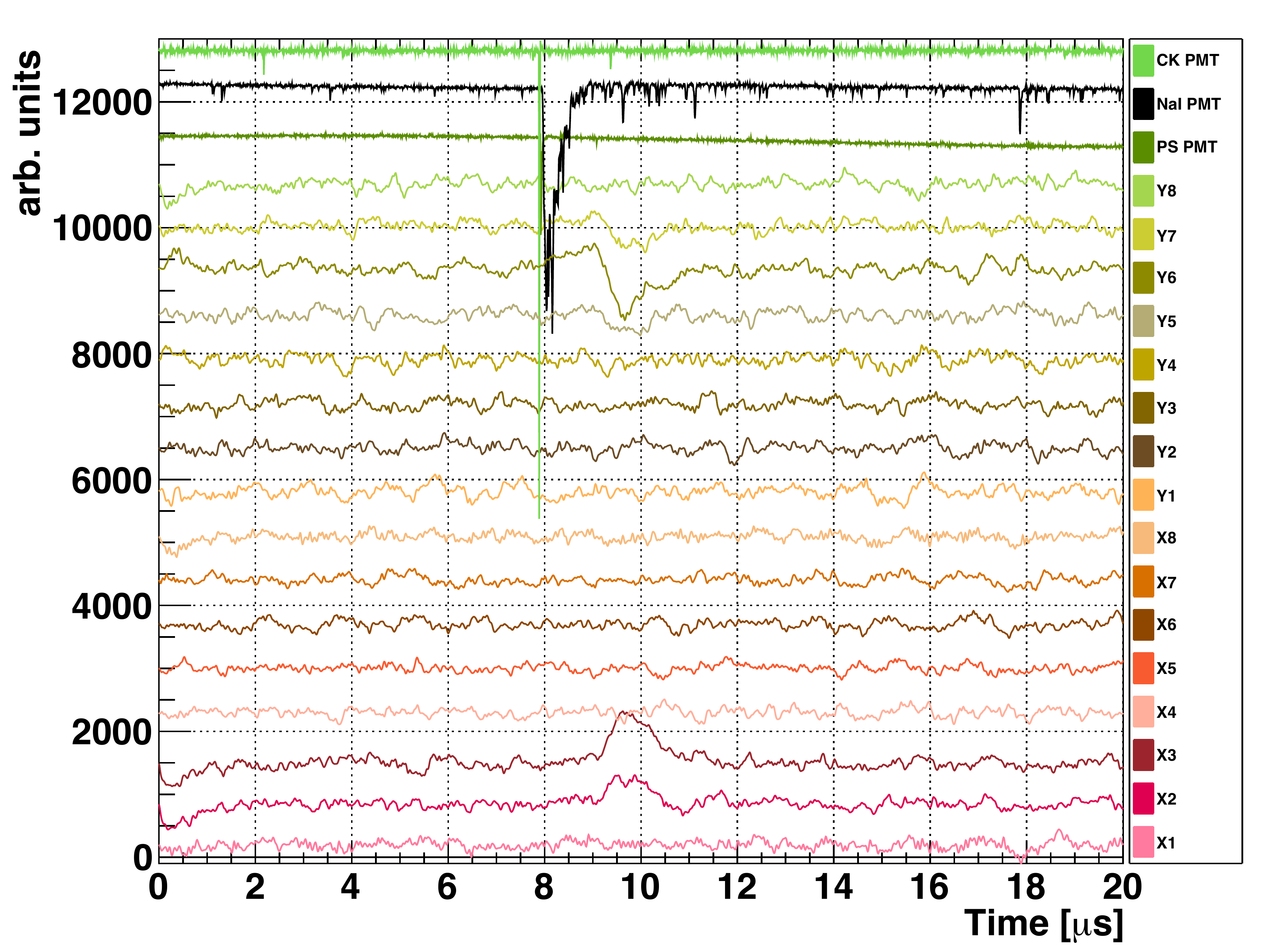}
\caption{A pair production candidate event. Simultaneous signals are observed in both the external tagging detectors and the Cherenkov PMT (denoted as ``CK PMT" in the figure), followed by hits in two $X$ and two $Y$ wires in the TPC. The wire and the PMT signals are offset for better visualisation.}
\label{fig:pd_event}
\end{figure}

\begin{figure}[ht!]
\includegraphics[width=0.95\linewidth]{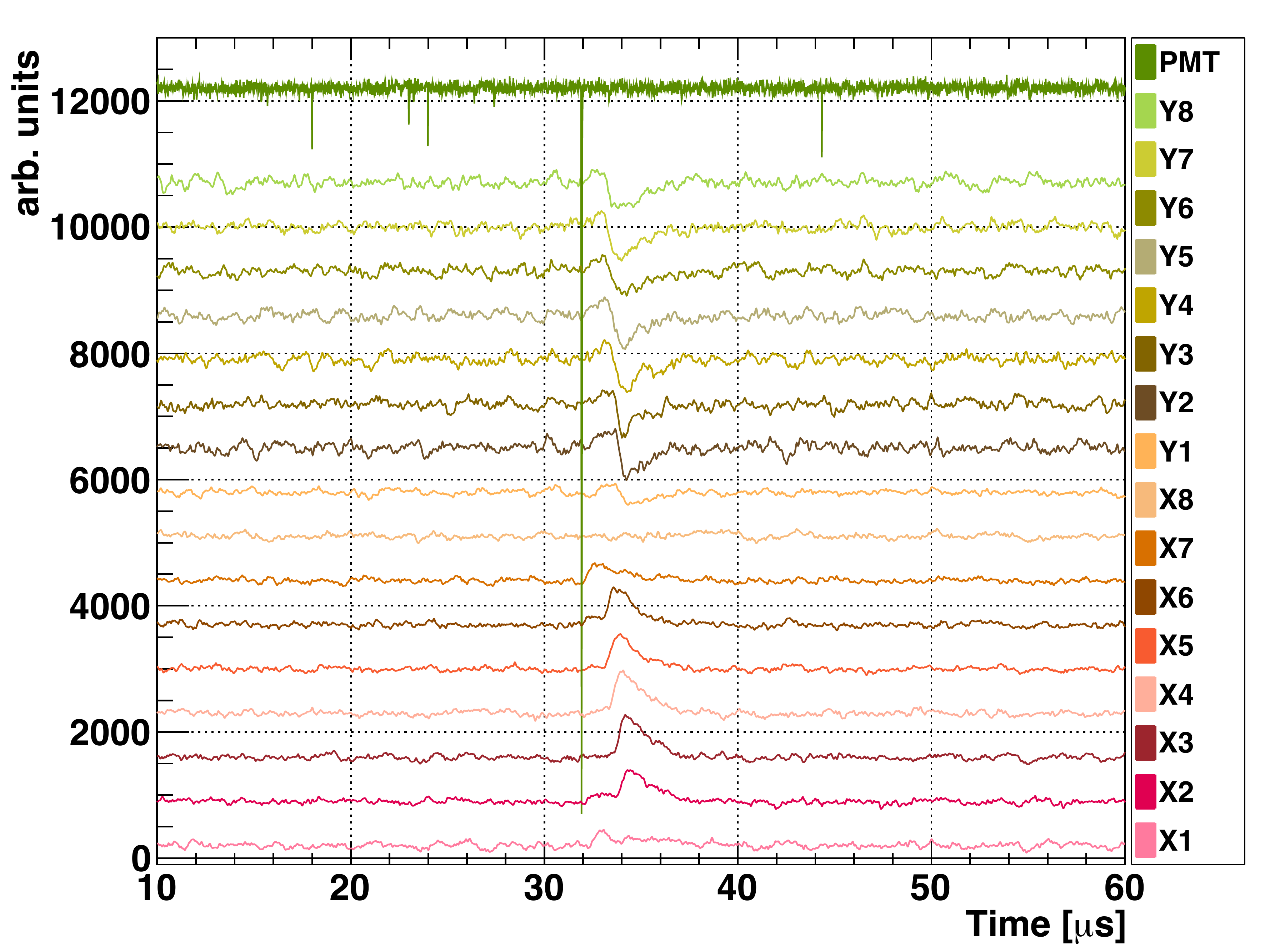}
\caption{Candidate event for a muon crossing the detector. Both the 8 $X$ and $Y$ wire signals are shown together with the Cherenkov PMT signal (denoted as ``PMT" in the figure). The trigger provided by the muon telescope (described in Section~\ref{sec:muons}) is happening at 32 $\mu$s. Signal traces are offset for better visualisation. Note the different time scale from Figure~\ref{fig:pd_event}.}
\label{fig:muon_event}
\end{figure}

The recorded binary data is processed through an analysis chain to decode the raw waveforms, then perform hit finding and track reconstruction.  An analysis threshold of $4\sigma$ above the baseline noise is applied to both the $X$ and $Y$ wire signals to define a ``hit" channel. The deposited energy is calculated from the raw waveforms by summing the charge detected by each of the $X$ wires which passes the threshold requirement.
We then apply digital filters to the waveforms, which improve timing and position reconstruction in the $Z$ direction. The integral and differential time constants are chosen to be 1 $\mu$s and 0.25 $
\mu$s. Wire hits are then defined with the following information associated: wire coordinate in $X$ or $Y$, drift time in $Z$ (calculated from the difference in time between the prompt trigger and the time at which the charge waveform exceeds the threshold) and amount of charge deposited. Examples of signal waveforms for all the 8+8 wires after applying the digital filters are shown in Fig.~\ref{fig:pd_event} (AmBe calibration) and Fig.~\ref{fig:muon_event} (muon calibration). Charge passes through the $Y$ plane producing bipolar induction signals, then is collected on the $X$ plane producing unipolar collection signals.

\section{Calibration with an AmBe $\gamma$-ray source}
\label{sec:ambe}

To calibrate the energy scale in the TPC, we use pair-production signals from 4.4~MeV $\gamma$-rays emitted by an AmBe source. Gamma rays are produced by the de-excitation of $^{12}$C nuclei following a $^{9}$Be$\left(\alpha,n\right)^{12}$C$^*$ reaction inside the source capsule. Pair production events in the TMS provide an ideal calibration as they are mono-energetic: energy is deposited in the form of an electron-positron pair with exactly $E_{\gamma}-2m_e$. The two $511$~keV $\gamma$-rays from the subsequent positron annihilation have a scattering length of 16~cm in TMS, and will therefore escape the active volume of the TPC with a high probability. By tagging the two annihilation gammas in a pair of secondary detectors, these events are identified with a high signal-to-noise ratio.

\subsection{Experimental setup and event selection}
\label{subsec:ambe_setup}

\begin{figure}[t!]
    \centering
    \includegraphics[width=1.\linewidth]{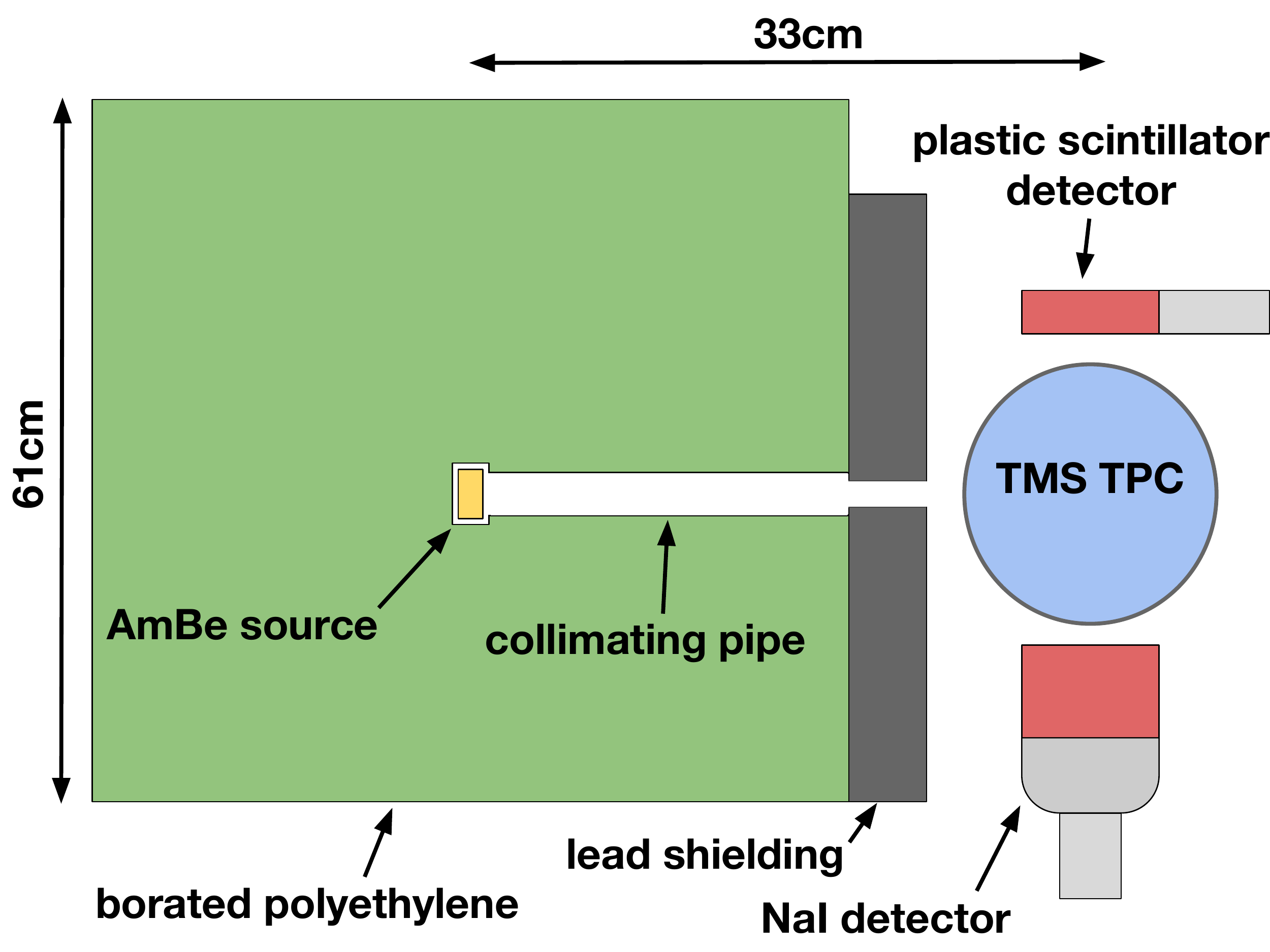}
    \caption{Illustration of the setup during the AmBe $\gamma$-ray calibration campaign. The plastic scintillator detector and NaI detector are placed back-to-back to tag the 511~keV positron annihilation $\gamma$-rays from pair production events in the TMS.}
    \label{fig:pair_production_setup}
\end{figure}

The setup is illustrated in Figure~\ref{fig:pair_production_setup}. The source\footnote{The activity of the $^{241}$Am in the source is $\sim$100~GBq, which produces approximately $10^6$ $(\alpha,n)$ reactions per second.} is placed in a shielding assembly made of boron-doped polyethylene (5\%-by-weight) to moderate and absorb the fast neutrons produced by the $(\alpha,n)$ reactions. A $\sim$5~cm diameter hole in the plastic shielding acts as a collimator for radiation in the direction of the TMS detector. An additional 5cm-thick lead collimator with a $2.5$cm$\times2.5$cm square opening is added at the end of the plastic shielding to block secondary $\gamma$-rays produced by neutron capture in the shielding material. To tag annihilation gammas from pair-production, two additional detectors are placed above and below the TPC: one EJ200 plastic scintillator (PS) detector (5~cm diameter $\times$ 10~cm length) and one NaI crystal (10~cm diameter $\times$ 10~cm length). The EJ200 detector provides a fast signal for precise coincidence tagging, and the NaI detector provides superior energy resolution, allowing the selection of events in which the coincident signals have energies consistent with the expected $511$~keV annihilation gammas.  Data acquisition is triggered by signals in the two tagging detectors arriving within a coincidence window of 100~ns.

The energy scale and timing properties of the tagging detectors are calibrated using $^{137}$Cs and $^{60}$Co $\gamma$-ray sources. To calibrate energy response in the NaI detector, the spectra around 662~keV, 1170~keV, and 1332~keV are used after fitting the peaks to gaussian distributions. For the plastic scintillator, the poor energy resolution prevents the observation of individual features. Instead, the compton edge observed in the $^{60}$Co spectrum is fitted to a gaussian-smeared step function to provide an approximate scaling. Timing coincidences are calibrated using the 1170~keV and 1332~keV $\gamma$-rays from the $^{60}$Co source, which are emitted approximately back-to-back. For the timing calibration, the apparatus is modified so that the detectors are positioned with the $^{60}$Co source directly in between, and data acquisition is triggered by pulses in the plastic scintillator. The measured coincidence peak is then fitted to a gaussian distribution to determine the timing offset (due to delays in the electronics) and the resolution. We measure a 1-$\sigma$ timing resolution of 5.8~ns between the plastic scintillator and the NaI detector, limited primarily by the relatively slow scintillation response of the NaI crystal. We also measure a timing resolution of 1.1~ns for coincidences between the plastic scintillator and Cherenkov light signals in the TMS. This is consistent with the transit time spreads of the photomultiplier tubes.

\begin{figure}
    \centering
    \includegraphics[width=0.95\linewidth]{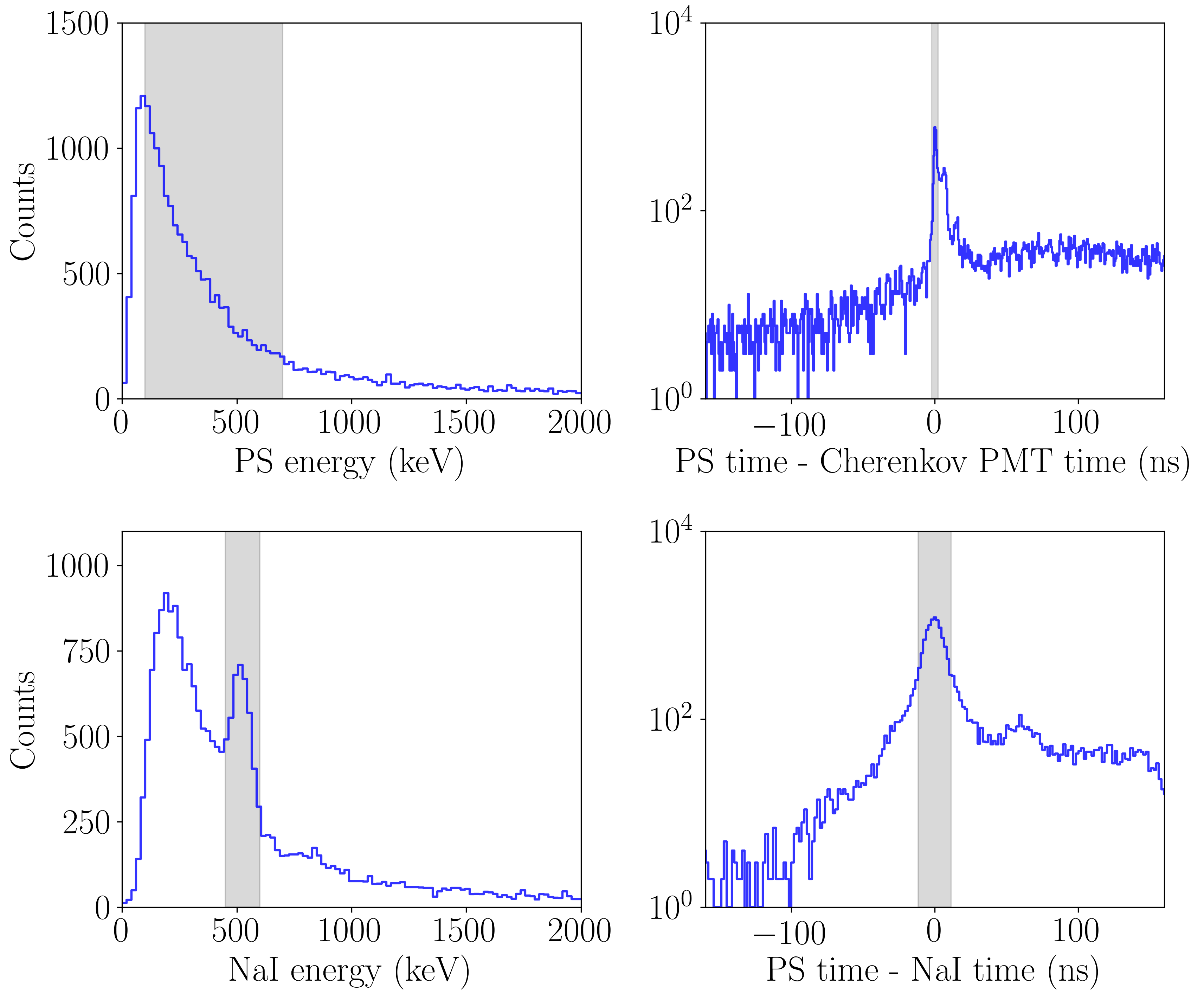}
    \caption{Energy (left) and timing (right) spectra provided by the external tagging detectors for events in the AmBe calibration campaign. The pair production selection cuts in energy and time are shown by the grey shaded regions. The energy (timing) spectra displayed here are shown after applying the timing (energy) cuts, to illustrate the distribution of events which are considered pair production candidates.}
    \label{fig:ambe_cuts}
\end{figure}

We select events in the AmBe calibration data using both energy and timing information from the two tagging detectors. Our selection cuts for pair production candidates are illustrated in Figure~\ref{fig:ambe_cuts}. Since  we are searching for 511~keV annihilation gammas, we require the energy in the PS detector to be between 100--700~keV (to account for poor energy resolution), and the energy in the NaI detector to be between 450--600~keV. We further require that the time differences between the NaI and PS detector and the PS detector and the Cherenkov PMT be within 2$\sigma$ of our $^{60}$Co-calibrated coincidence time. Events passing all four selection criteria are considered pair-production candidates. The background is evaluated by studying the sidebands for each cut, and is estimated to contribute less than 6\% of the selected events.

Due to the tagging detectors' vertical orientation, our data acquisition is frequently triggered by downward-going cosmic ray muons. As muons are expected to deposit several MeV in a few cm of material, the scintillation signals in the external detectors are large enough to saturate the data acquisition. We can therefore select downward-going muon events with high efficiency by requiring that both external detectors saturate in a single event.

\subsection{Cherenkov light response}
\label{subsec:cherenkov}

The Cherenkov light produced in the TMS provides a prompt signal that could be used for self-triggering. The refractive index in TMS at 20$^{\circ}$C is $n=1.359$~\cite{TMSRefractiveIndex}, resulting in Cherenkov thresholds for electrons and muons of 289~keV and 50.5~MeV, respectively. To identify Cherenkov light signals in the TPC, we compare three classes of events: uncorrelated events (where we expect only random coincidences and noise in the Cherenkov PMT), pair-production candidate events, and cosmic ray muon candidate events. Pair production candidates and muon candidates are selected as described in Section~\ref{subsec:ambe_setup}. The uncorrelated event sample is generated by selecting Cherenkov pulses arriving $>80$\,ns before the calibrated coincidence trigger time, well outside of the physical coincidence window. For these events, we also do not apply any coincidence or energy selection criteria in the tagging detectors, in order to get an unbiased sample of background noise.

\begin{figure}
    \centering
    \includegraphics[width=1.\linewidth]{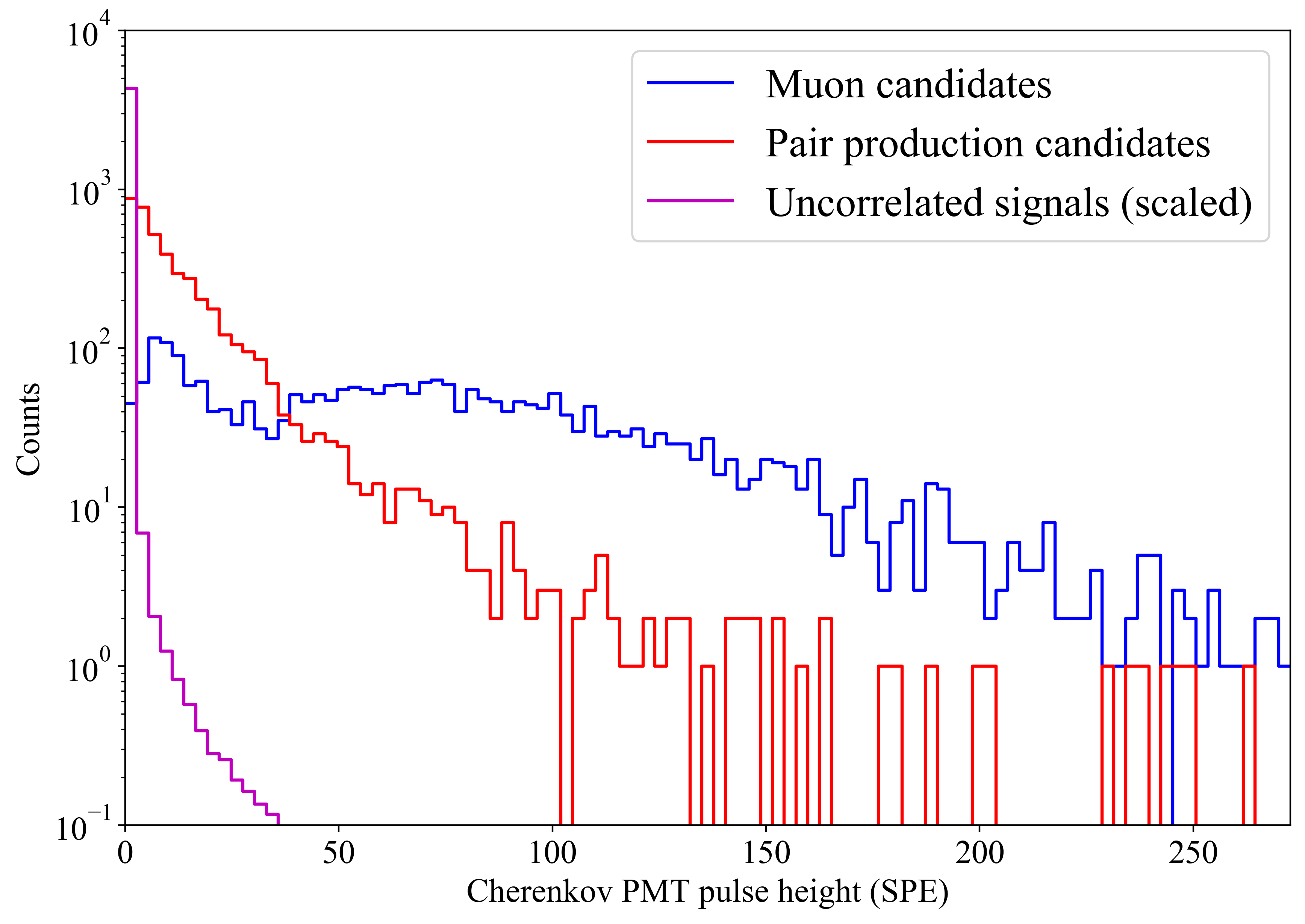}
    \caption{Pulse height distribution recorded by the PMT reading out the Cherenkov light in TMS for cosmic ray muons and 3.4~MeV fast electrons, compared with uncorrelated noise. For direct comparison, the noise distribution is scaled so that its integral matches the number of events passing the pair production selection cuts. Cherenkov signals for both fast electrons and muons are observed well above the uncorrelated noise, providing a signal that can be used for self-triggering in a LOr TPC.
    }
    \label{fig:cherenkov_light}
\end{figure}

Figure~\ref{fig:cherenkov_light} shows the distribution of observed pulse heights recorded by the Cherenkov PMT. A Cherenkov signal is clearly visible above background for both fast electrons and muons. A quantitative analysis of the Cherenkov light spectra and tagging efficiency requires a complete optical model of the detector internals and a model of light attenuation in the optical coupling components, and is left for future work. However, this observation opens up the possibility of full 3D event reconstruction in a LOr TPC. 

\subsection{Charge calibration with pair-production events}
\label{subsec:pair_production}

Pair production in the TMS detector deposits a total of 3.4~MeV, which we use to calibrate the energy scale and energy resolution of the TMS detector. The average track length for electron/positron pairs at 3.4~MeV in TMS is $\sim$20mm (estimated using a Geant4~\cite{Geant4_1,Geant4_2} particle transport simulation). We therefore require hits on 2 or more wires in both $X$ and $Y$. The total charge is then defined as the sum of the raw charge signals on all $X$ wires that are hit. One candidate pair production event, with hits on two $X$ and two $Y$ wires, is illutrated in Fig.~\ref{fig:pd_event}.

\begin{figure}[t!]
\centering
\includegraphics[width=1.\linewidth]{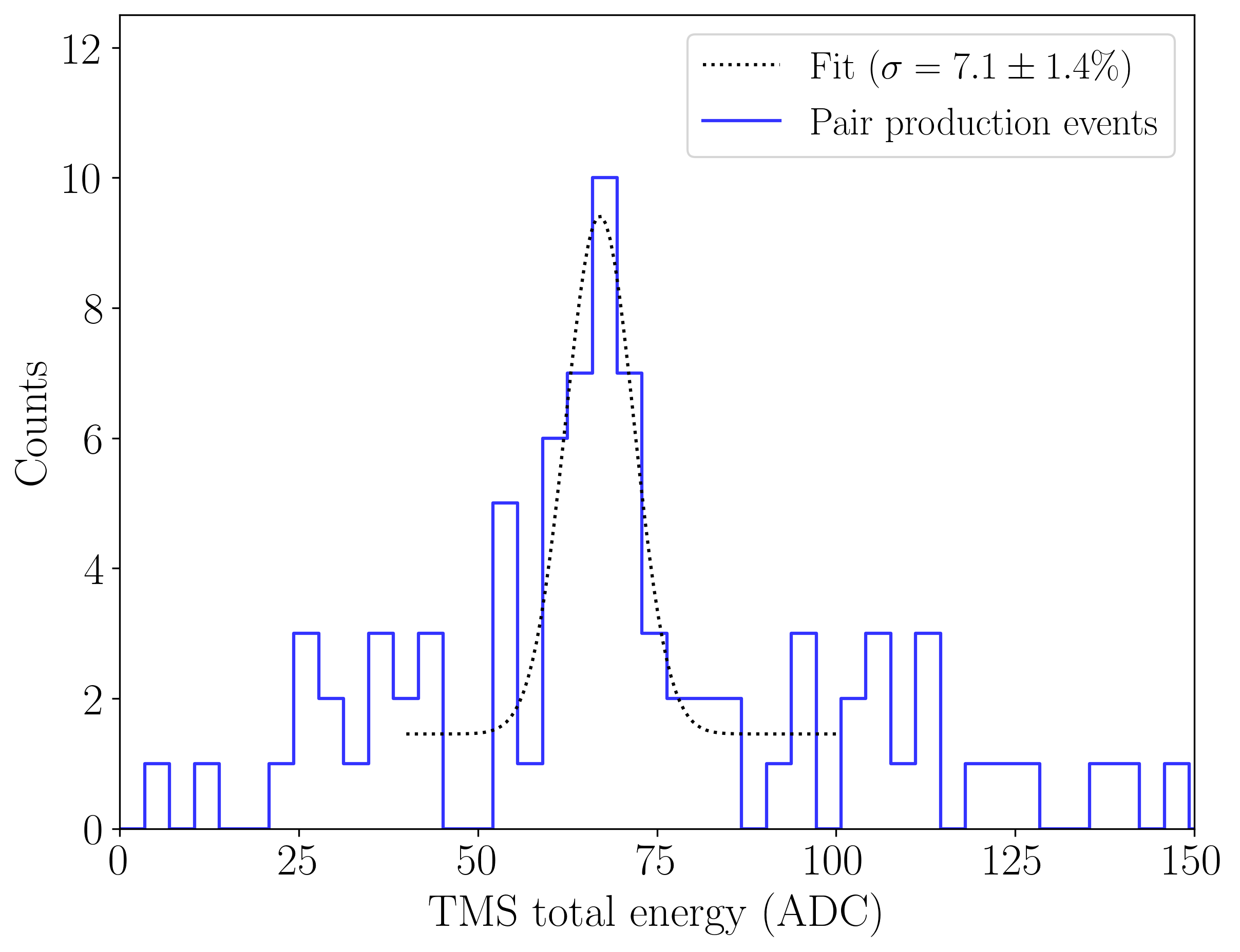}
\caption{Measured charge spectrum for pair production by 4.4~MeV gamma rays in the TMS detector, using approximately four live-days of data. We measure an energy resolution of $\sigma=(7.1\pm1.4)\%$ at 3.4~MeV, where the reported uncertainty is the statistical uncertainty in the fit.}
\label{fig:pp_spectrum}
\end{figure}

The resulting ionization spectrum is shown in Figure~\ref{fig:pp_spectrum}, using four days of acquired data. We fit the peak to a normal distribution with a constant background, finding a calibration constant of $19.6\pm1.3$~ADC/MeV (stat.). The width of the fit gives a 1$\sigma$ energy resolution of ($7.1\pm1.4)\%$ (stat.). The systematic uncertainty, due primarily to our choice of wire hit threshold, is estimated by shifting the threshold and re-fitting the spectrum, and is found to be $<$5\% of the measured quantity and hence subdominant.

\section{Calibration with cosmic ray muon tracks}
\label{sec:muons}

\begin{figure}[t!]
\includegraphics[width=0.54\linewidth]{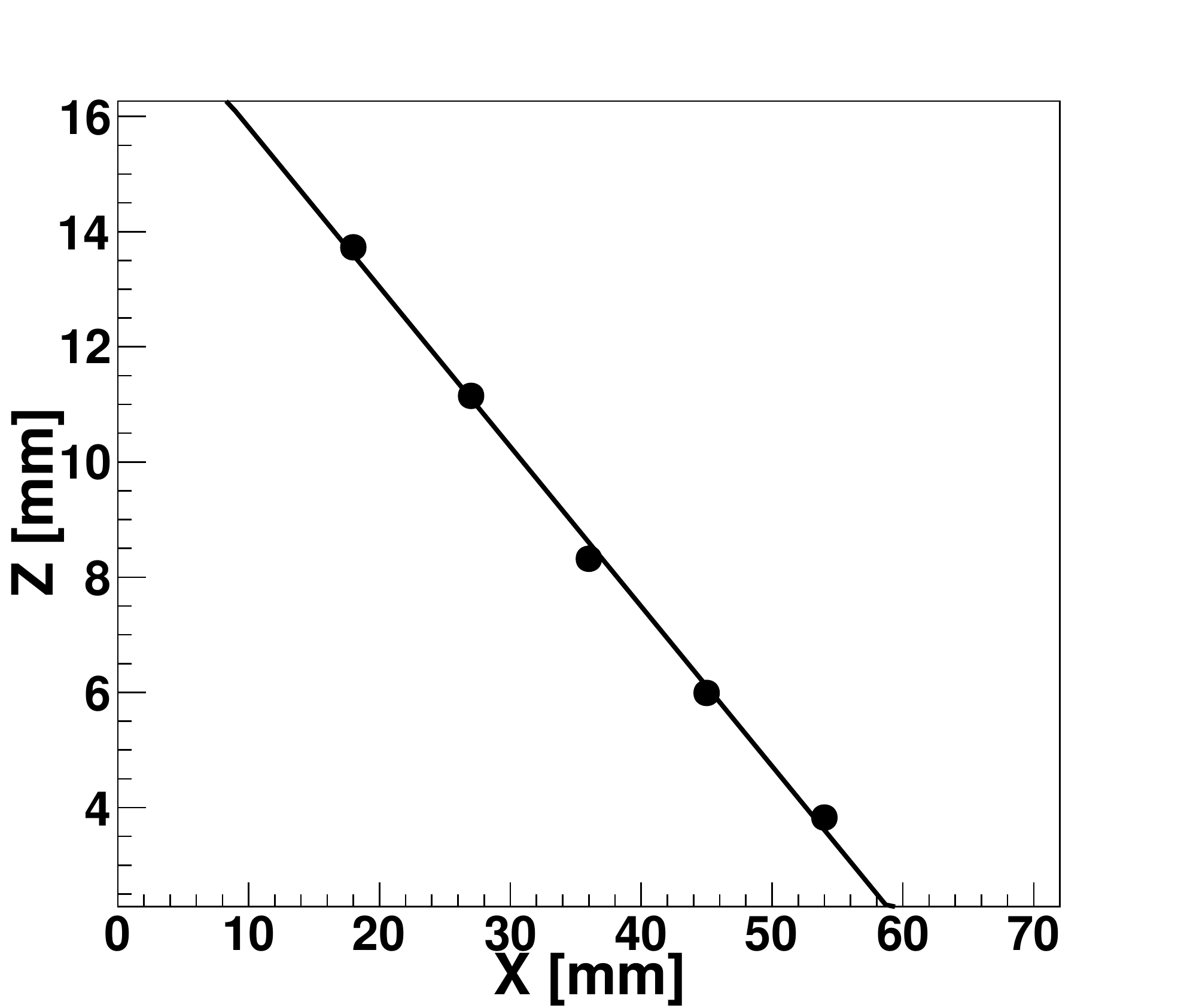}\includegraphics[width=0.54\linewidth]{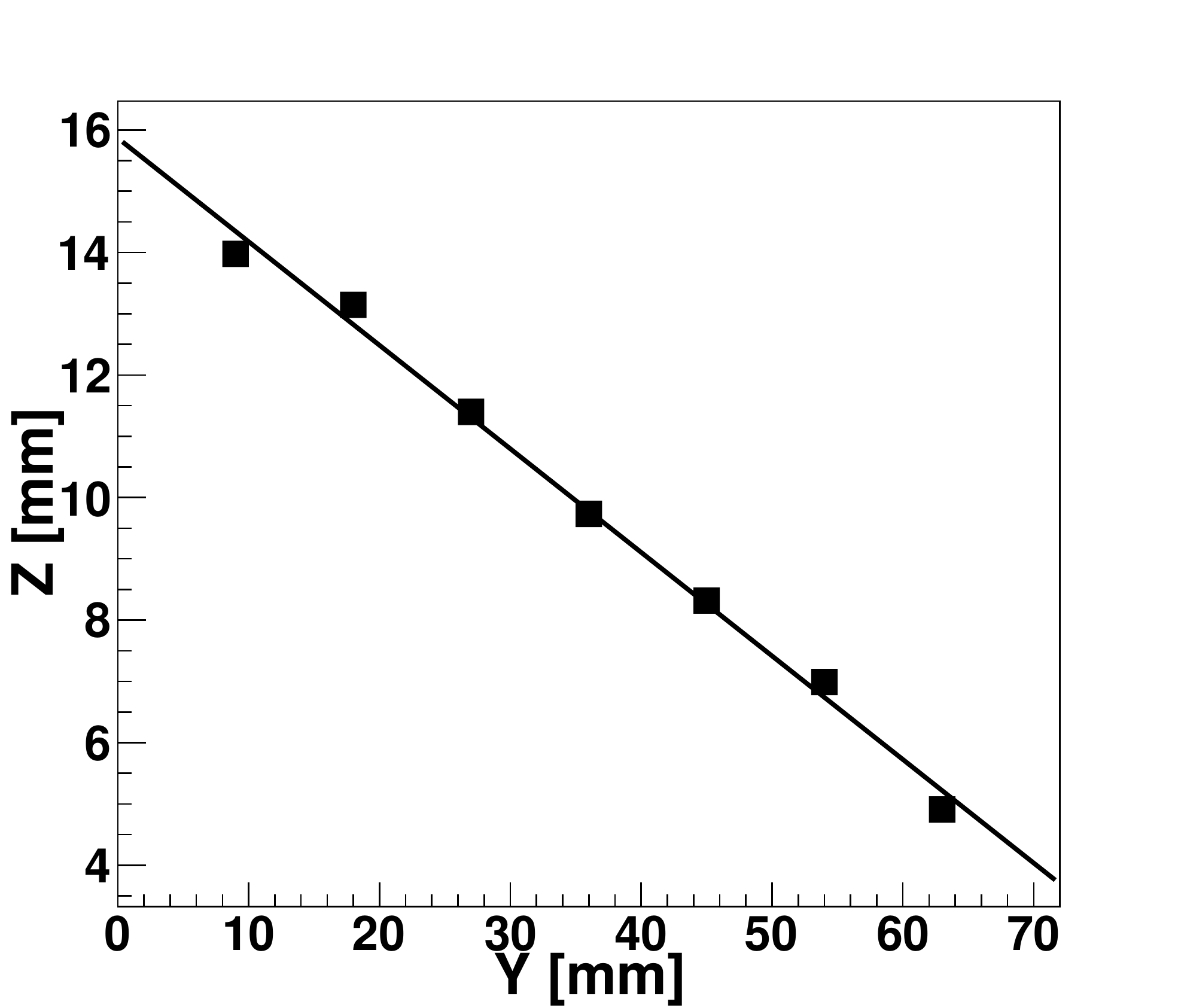}
\centering\includegraphics[width=0.54\linewidth]{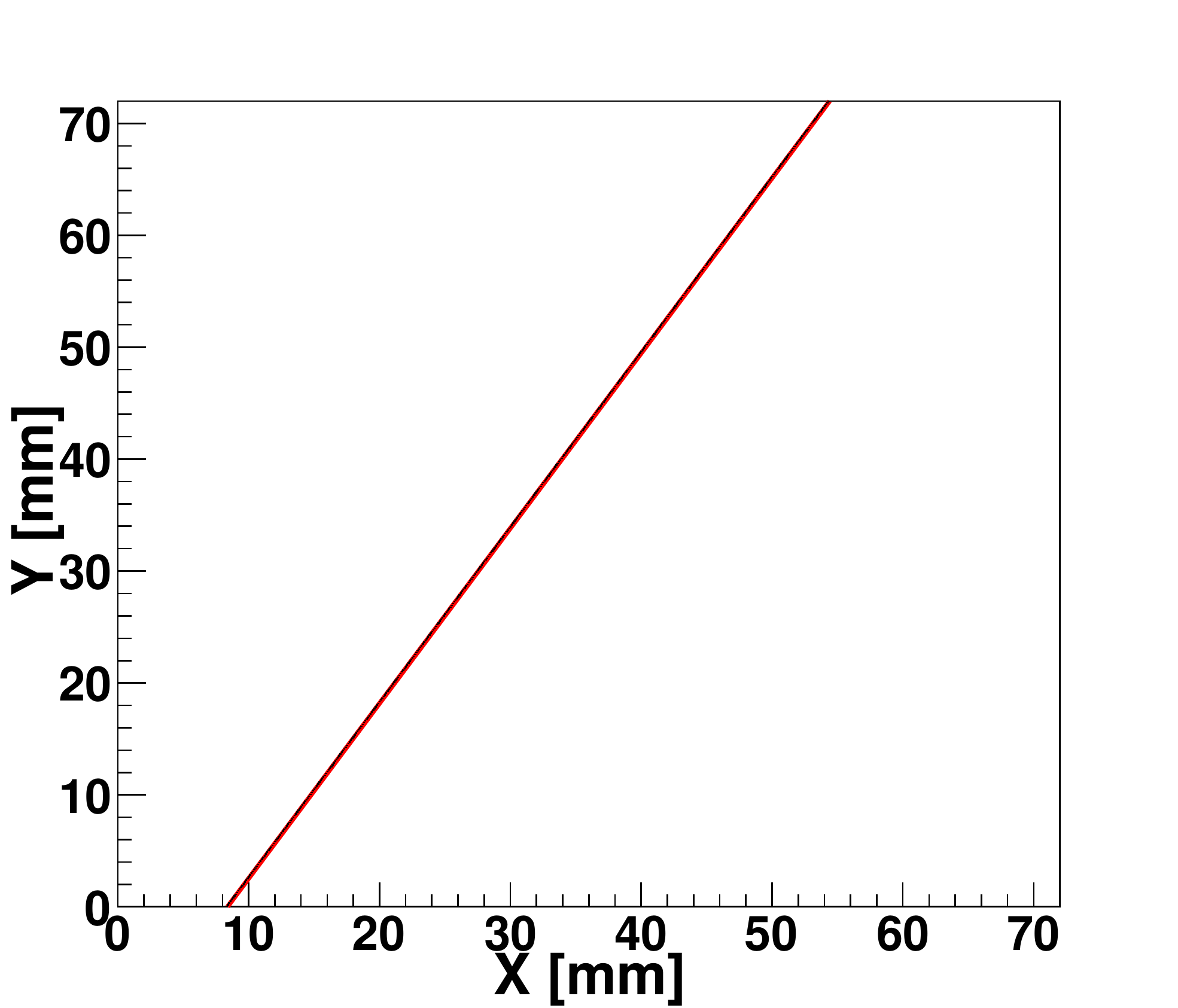}
\caption{An example of 3D track reconstruction. Top left (right): the $Z$-$X$ ($Z$-$Y$) spatial coordinates together with a linear fit; bottom: the matched $X$-$Y$ correspondence.}
\label{fig:3d_track_reco}
\end{figure}

We next perform measurements of the energy loss from minimum ionizing particles (MIPs) and the electron lifetime in the TMS, by reconstructing cosmic ray muon tracks in the detector. For these measurements, we replace the NaI crystal with a second EJ200 PS scintillator, and trigger on events which saturate both tagging detectors within a 100~ns coincidence window. For muon tracks, the number of hits exceeding the amplitude requirement is set to be $\geq$2 ($\geq$6) for the $X$ ($Y$) wires. This difference is due to the orientation of the two grids with respect to the vertical line connecting the two external detectors. Hits are identified with associated $X$ or $Y$ coordinates. The Z coordinate is calculated from the product of the known electron drift velocity and the measured drift time, where the latter is defined as the difference between the prompt trigger and the time at which the charge waveform exceeds the threshold. 
The maximum drift time is determined to be $3.0\pm0.1$~$\mu$s, resulting in a drift velocity of (5.2$\pm$0.2)~mm/$\mu$s, in a good agreement with the calculation based on the mobility~\cite{Engler_1996}. Two linear fits to the selected hits are applied to both the $Z$-$X$ and $Z$-$Y$ spatial coordinates to reconstruct the 3D track projections onto the two wire planes. After reconstructing the projections, the $X$-$Y$ correspondence is determined by the $Z$ coordinates in common. The 3D track reconstruction procedure is illustrated in Fig.~\ref{fig:3d_track_reco}. In addition to making a selection on the goodness of the linear fits, only tracks that cross the fiducial volume of the detector are retained by applying a selection criteria on the endpoints of the reconstructed 3D tracks. 

Tracks reconstructed in this way are then used to measure the stopping power and the electron lifetime.

\subsection{Stopping power $\bf{dE/ds}$}
\label{stopping_power}

\begin{figure}[t!]
\includegraphics[width=1.0\linewidth]{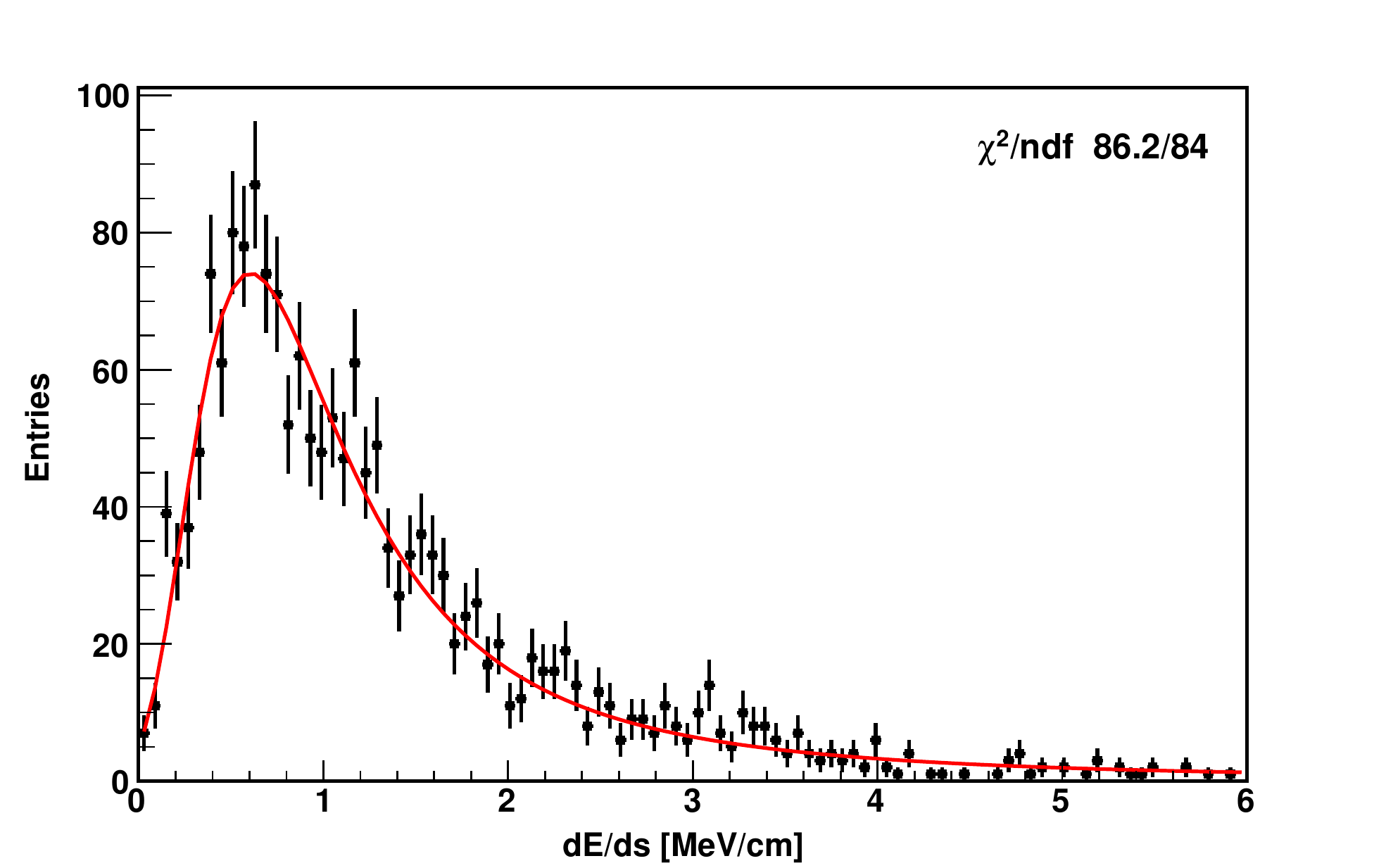}
\caption{$dE/ds$ measured for MIPs inside TMS together with the fit to a Landau convoluted with a Gaussian functions, in red.}
\label{fig:dedx_mip}
\end{figure}

Having reconstructed tracks in 3D, the charge $dQ$ collected for each track segment $ds$ can be extracted. Only $dQ$s from the $X$ wires are used since the charge is fully collected. The energy deposit $dE$ for a track segment $ds$ can be derived from the calibration constant of 20~ADC/MeV in Section\ref{subsec:pair_production}. The energy deposit $dE$ is further corrected against the charge loss due to the finite electron lifetime to be described in Section~4.2, although for the present data, this correction is negligible. The $dE/ds$ distribution is shown in Fig.~\ref{fig:dedx_mip}. The distribution is well described by the Landau function~\cite{Landau} convoluted with a Gaussian function to describe the instrumental resolution shown as the red curve. In the fit, all the 4 parameters, width of the Landau distribution, most probable value, integral and sigma of the Gaussian function, are set to float. We obtain the most probable energy loss per unit length from the fit: $\Delta_p/ds$ = (0.60$\pm$0.01) MeV/cm, where the error is statistical, resulting from the fit. The systematic error for the $\Delta_p/ds$ measurement is studied by changing the binning and fitting range, resulting in a $\pm2\%$ uncertainty of the best fit value. The resolution of the $\Delta_p/ds$ measurement defined as the sigma of the Gaussian divided by the $\Delta_p/ds$ is 4.7\%. This is in good agreement with the value obtained from the gamma pair production calibration accounting for the energy deposited on each wire. 

A mean energy loss of 1.29~MeV/cm for MIPs inside TMS is obtained by averaging the $dE/ds$ distribution shown in Fig.~\ref{fig:dedx_mip} and is consistent with a value of 1.3~MeV/cm obtained by using the the nominal stopping power of 2 MeV$\cdot$cm$^2$g$^{-1}$ for MIPs~\cite{PDG} and a TMS density of 0.65 g/cm$^3$~\cite{Engler_1996}.

\subsection{Electron lifetime}
\label{lifetime}

\begin{figure}[t!]
\includegraphics[width=1.0\linewidth]{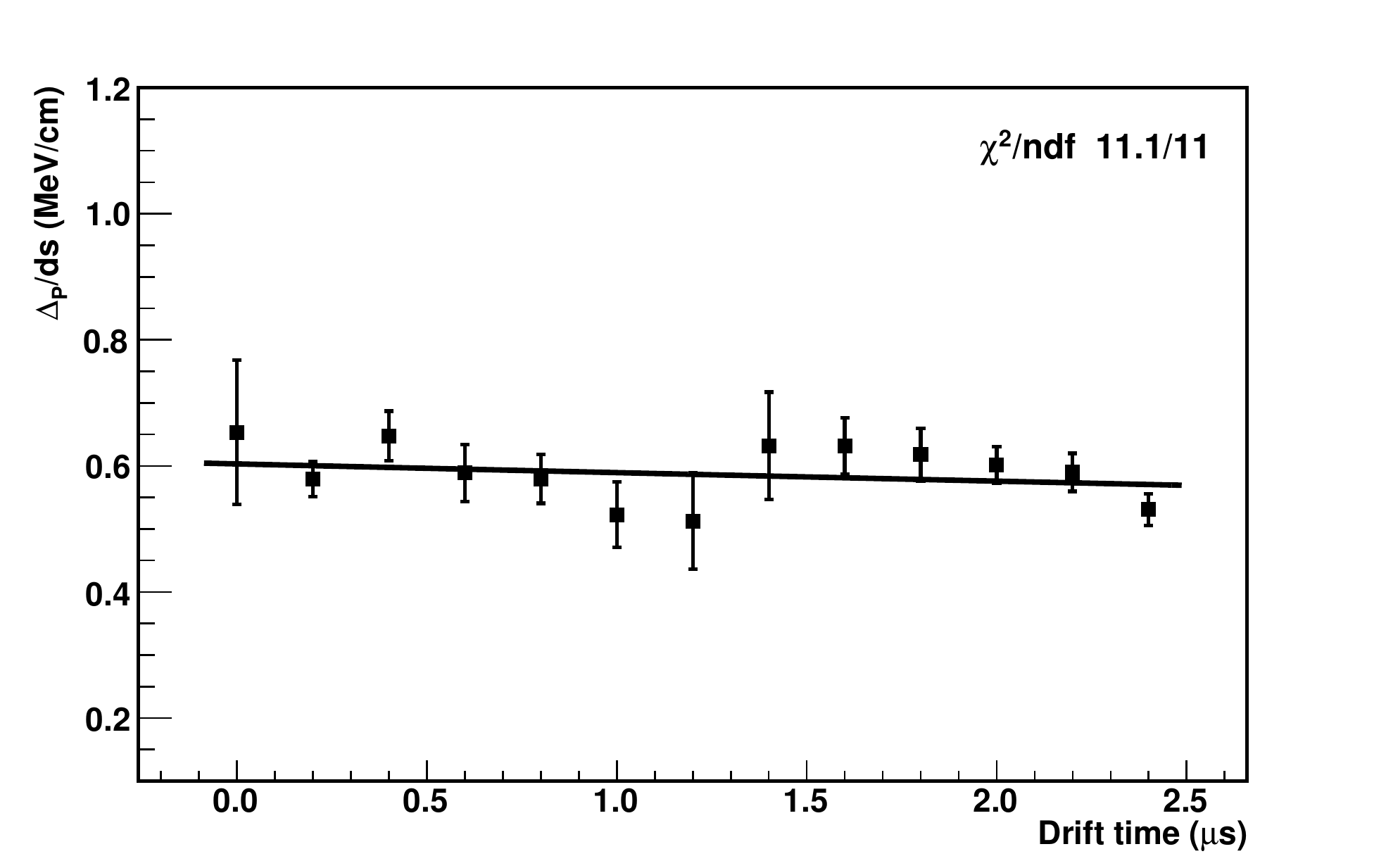}
\caption{Most probable energy loss as a function of drift time along with an exponential fit to the data points.}
\label{fig:dqdx_dt}
\end{figure}

The 3D readout of the TPC allows for a proper measurement of the electron lifetime. $dE/ds$ distributions without purity correction can be obtained for bins of different drift time. For each such distribution, the same procedure of fitting described in the previous section is applied to extract the most probable energy losses as per unit length $\Delta_p/ds$, shown in Fig.~\ref{fig:dqdx_dt}. It is clear that the electron lifetime $\tau_e$ obtained is too long to result in a significant attenuation in the short drift distance used. Nevertheless, an exponential fit can be used mainly to provide a lower bound on $\tau_e$.
This procedure obtains $\tau_e=$43$^{+680}_{-21}$~$\mu$s, consistent with the values between 
10 and 100 $\mu$s reported in previous measurements~\cite{engler_1986,Faidas_1990,Lopes_1988,Ochsenbein_1988}. Systematic effects are estimated by changing the fit range and binning, and are found to be subdominant.
To our knowledge, this is the first case in which the measurement of the charge yield and the detector lifetime in a LOr are clearly decoupled.

\section{Conclusion}

This work describes the design, operation, and first measurements of a TPC filled with tetramethylsilane. Using a PMT coupled to the liquid volume, we observe coincident Cherenkov signals for fast electrons and muons which could be used as a self-trigger without the need of an external trigger. Three dimensional tracking is demonstrated, albeit in a small detector. Charge yield and electron lifetime in TMS are independently measured. Given the short drift distance, the electron attenuation is negligible and an exponential fit returns a lifetime of 43$^{+680}_{-21}$~$\mu$s. Future work with a longer detector will improve this measurement. 

A self triggering TPC filled with organic liquids may hold future promise in advanced detectors for antineutrinos, medical imaging, or other applications requiring large scale radiation detectors with precise position resolution.

\section{Acknowledgements}
We would like to thank L.~Fabris (ORNL) and R.~G.~DeVoe (Stanford) for discussions and help on the charge readout electronics. S.~W. and B.~L. acknowledge the partial support of the Karl A. Van Bibber Postdoctoral Fellowship Endowment from Stanford physics department. This work was supported by seed funds of Stanford University. 

\bibliographystyle{plain}
\bibliography{tms_bibliography}
\end{document}